\documentclass[pra,twocolumn,superscriptaddress,showpacs,preprintnumbers,amsmath,amssymb]{revtex4}
\usepackage{amsmath,amssymb}
\usepackage{color}
\usepackage{epsfig}
\usepackage{dcolumn}
\usepackage{bm}

\newcommand{\comment}[1]{}

\newcommand{\beq}{\begin{eqnarray}}
\newcommand{\eeq}{\end{eqnarray}}

\newcommand{\lan}{\langle}
\newcommand{\ran}{\rangle}

\input xy 
\xyoption{all}

\begin{document}

\bibliographystyle{elsart-num}

\title{Inadequacy of a classical interpretation of quantum projective measurements via Wigner functions}

\author{Amir Kalev}
\affiliation{Centre for Quantum Technologies, National University of Singapore, Singapore 117543\footnote{Present address.}}
\affiliation{Department of Physics, Technion-Israel Institute of Technology, Haifa 32000, Israel}

\author{Ady Mann}
\affiliation{Department of Physics, Technion-Israel Institute of Technology, Haifa 32000, Israel}

\author{Pier A. Mello}
\affiliation
{Instituto de F\'isica, Universidad Nacional Aut\'onoma de M\'exico, Ap. Postal 20-364, 01000 M\'exico, D. F., Mexico}

\author{Michael Revzen}
\affiliation{Department of Physics, Technion-Israel Institute of Technology, Haifa 32000, Israel}

\date{\today}

\begin{abstract}
We study the possibility of giving a classical interpretation to quantum projective measurements for
a particle described by a pure Gaussian state whose Wigner function is non-negative. 
We analyze the case of a projective measurement which gives rise to a proper Wigner function, i.e., taking on, as its values, the eigenvalues of the projector. We find that, despite having this property, this kind of projector produces a state whose Wigner function ceases to be non-negative and hence precludes its interpretation as a classical probability density.
We also study the general case in which the projected state has a non-negative Wigner function; but then we find that the Wigner function of the projector is not a proper one.
Thus, we conclude that a non-negative Wigner function is inadequate to serve as a hidden variable model for quantum processes in which projective measurements take place.
\end{abstract}


\maketitle


It is well known that for a system possessing continuous dynamical variables Wigner functions (WFs) allow computing quantum-mechanical expectation values as integrals over phase space \cite{schleich}.
However, the WF of a quantum-mechanical state is not necessarily non-negative, thus precluding, in general, its interpretation as a classical probability density in phase space.
In contrast, in those cases when the WF of a state {\em is} non-negative, one is tempted to interpret the phase space variables $q, p$ as hidden variables (HVs) having ``physical reality'', endowed with a ``classical" probability density given by the WF.
However, Ref. \cite{rev} presented a detailed analysis of the problem of a Bell pair of particles described by Gaussian wavefunctions (recall that for such wavefunctions the WF is non-negative). 
That analysis showed that non-negativity of the WF of the states is not sufficient to enable one to associate a local hidden variable (LHV) model with quantum mechanical expectation values. 
A further condition is required and should be imposed on the observables involved. In the terminology of Ref. \cite{rev}, the WFs of the observables must
qualify as ``proper observables in phase space":
by this we mean that the WF of an observable $\hat{A}$ must take on, as its values, precisely the eigenvalues of the operator $\hat{A}$.
A simple example which will be useful below is the observable 
[considered in Eq. (2.16) of Ref. \cite{rev}]
\begin{subequations}
\begin{eqnarray}
\hat{A}
=\int_{-\infty}^{\infty}|x\ran  f(x) \lan x|\, dx\;,
\label{A}
\end{eqnarray}
defined through its spectral representation, whose eigenvalues are $f(x)$, and whose WF
\begin{eqnarray}
W_{\hat{A}}(q,p)
=f(q)
\label{W_A}
\end{eqnarray}
\label{A,W_A}
\end{subequations}
takes on, as its values, precisely the eigenvalues of the operator $\hat{A}$.
Gaussian wave functions are also widely employed in the problem of continuous-variable teleportation (a three-particle problem);
in this context it is sometimes stated that one can describe the problem in terms of LHVs endowed with a classical probability density
\cite{caves_wodkiewicz,kalev_2007}.

In the present Brief Report we contemplate the problem of quantum-mechanical projective measurements and analyze the possibility of interpreting them according to the rules of classical statistics when, before the measurement, we have a pure state described by a non-negative WF.
For simplicity and in order to be as concrete as possible we concentrate on a single-particle system.
We first study the problem of a particular selective projective measurement associated with position. We then undertake the analysis of a more general case.


Consider a one-particle system described by the state 
$|\psi \ran$ and
suppose we make a measurement of position which consists in selecting the portion $(-a/2, a/2)$ of the $x$-axis.
The resulting state is obtained by applying to $|\psi \ran$ the projector
\beq
\hat{P}
=\int_{-a/2}^{a/2}|x\ran\lan x|\, dx\;,
\label{P}
\eeq
thus giving
\beq
|\psi'\ran
=\frac{\hat{P}|\psi\ran}{\sqrt{\lan\psi|\hat{P}|\psi\ran}}
=\frac{\int_{-a/2}^{a/2}\psi (x) \,| x \ran \, dx}
{\sqrt{\int_{-a/2}^{a/2}|\psi (x)|^2\, dx}}\;.
\label{psi_(-a/2,a/2)}
\eeq
The projector of Eq. (\ref{P}) has two eigenvalues: $\{1,0\}$. 
The result (\ref{psi_(-a/2,a/2)}) of the measurement corresponds to having selected the subensemble with the eigenvalue $1$.
In this sense, we shall refer to this operation as a ``selective projective
measurement" \cite{lars2007}.

We now wish to describe the projection process in terms of WFs.
The WF of a pure quantum state $|\psi\ran$ for a one-particle system is given by (this definition can be extended to a mixed state, as well as to a multi-particle system) \cite{schleich}
\beq
W_{|\psi\ran}(q, p)
= \int_{-\infty}^{\infty}e^{-ipy}\lan q + y/2| \psi\ran\lan\psi |q - y/2 \ran dy \; ,
\label{WTrho}
\eeq
for which we adopt the normalization condition
\beq
\int \int_{-\infty}^{\infty} W_{|\psi\ran} (q,p) \; dq \; \frac{dp}{2 \pi} =1 .
\label{norm}
\eeq

If the original state {\em before the measurement} is the pure Gaussian state
\beq
|\psi\ran=\int_{-\infty}^{\infty} dx \frac{e^{-\frac{x^2}{4\sigma^2}}}{(2\pi\sigma^2)^{1/4}} |x\ran\;,
\label{psiGauss}
\eeq
its WF is given by
\beq
W_{|\psi\rangle} (q,p)
=2\,e^{-\frac{q^2}{2\sigma^2}}\,e^{-2\sigma^2p^2}
\label{WPsiGauss}
\eeq
and is non-negative.
The WF of the projector of Eq. (\ref{P}), which is our observable in this case, is given by
\beq
W_{\hat{P}}(q,p)
=
\theta \left(\frac{a}{2}-|q| \right),
\label{W_P}
\eeq
(where $\theta(x)=1,0$ for $x \geq 0$, $x < 0$, respectively, is the usual step function),
and takes on the values $1$ and $0$, the eigenvalues of the projector.
This is a particular case of Eqs. (\ref{A,W_A}) above.
These facts suggest that we may interpret the variables 
$q, p$
as HVs, $W_{|\psi\rangle}$ playing the role of a probability density in phase space.

\begin{figure}[ht]
\epsfxsize=.50\textwidth
\epsfysize=.50\textwidth  
\centerline{\epsffile{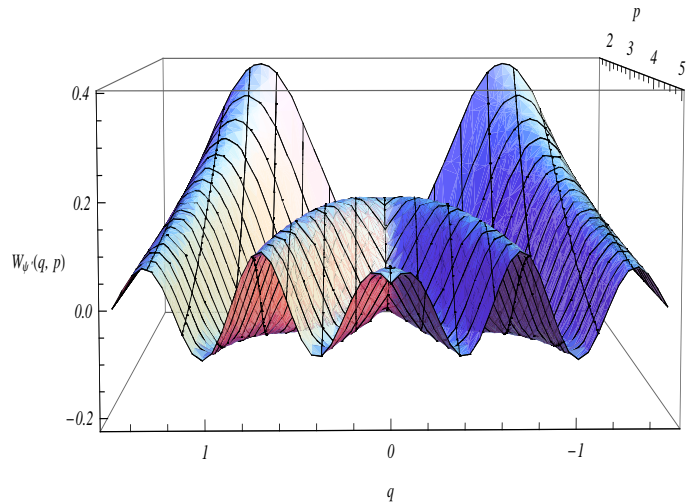}}
\caption{The WF $W_{|\psi'\ran} (q, p)$ of the projected state $|\psi'\ran$ after a selective projective measurement, given in Eq. (\ref{W_psi_(-a/2,a/2)}) and evaluated for $\sigma =1$ and $a=3$; 
it is plotted in the interval $-1.5 \leq q \leq 1.5$ 
($W_{|\psi'\ran} (q, p) = 0$ outside this interval), and 
$1.5 \leq p\leq 5$. 
Notice that this WF is not non-negative.}
\label{fig1}
\end{figure}

{\em After the measurement}, the WF of the projected state 
$|\psi'\ran$ of Eq. (\ref{psi_(-a/2,a/2)}) is given by
\begin{eqnarray}
W_{|\psi'\ran} (q, p)
&=&\frac{W_{\hat{P}|\psi\ran}(q,p)}{\lan\psi|\hat{P}|\psi\ran}
\nonumber \\
&=&\frac{\Re[\textrm{erf}(z_1)]}{\textrm{erf}(\frac{a}{2\sqrt{2}\sigma})} \; 
W_{|\psi\ran}(q,p) \theta\left(\frac{a}{2}-|q|\right) ,
\label{W_psi_(-a/2,a/2)}
\end{eqnarray}
where $\textrm{erf}(z)= \frac{2}{\sqrt{\pi}}\int_0^z e^{-t^2} dt$
is the error function (Ref. \cite{abramowitz}, p. 297) and
\begin{equation}
z_1=\frac{\frac{a}{2}-|q|}{\sqrt{2}\sigma}+i\sqrt{2}\sigma p .
\label{z}
\end{equation}
The result given in Eq. (\ref{W_psi_(-a/2,a/2)}) has the important feature of becoming negative for certain values of $q$ and $p$.
This is not a surprise, because the state of Eq. (\ref{psi_(-a/2,a/2)})
in the coordinate representation is a non-Gaussian pure state, and by Hudson's theorem \cite{hudson} its WF cannot be non-negative. 
The possible negativity of its WF is shown in Fig. \ref{fig1} for the values of the parameters indicated in the figures.
Thus $W_{|\psi'\ran} (q, p)$ is not interpretable as a HV density.

In conclusion, {\em if we represent the above quantum mechanical projection via WFs, we cannot interpret the result in terms of a HV model endowed with classical statistical properties.}

We warn the reader against finding the WF of the projected state from the original $W_{|\psi\ran}(q,p)$ by just selecting in phase space the region $|q| < a/2$ 
(and renormalizing the result), as one would do
in classical statistics, because he would find the wrong answer. 
Indeed, the WF $W_{\hat{P} \hat{\rho} \hat{P}}(q,p)$ cannot be written as 
$W_{\hat{P}}(q,p)W_{\hat{\rho}}(q,p)$.
Notice also that, in Eq. (\ref{WPsiGauss}), $(\Delta q)^2=\sigma^2$ and $(\Delta p)^2=1/4\sigma^2$, with $(\Delta q)^2 (\Delta p)^2 =1/4$, the minimum value allowed by the uncertainty principle; selecting only the region $|q| < a/2$ would decrease $(\Delta q)^2$ without altering $(\Delta p)^2$, thus violating the uncertainty principle.

It is also instructive to illustrate the conflict between a quantum projection and a HV model via the expectation value of an observable with a proper WF. 
Consider the expectation value 
\beq
\lan\psi|\hat{P}\hat{A}\hat{P}|\psi\ran\;,
\label{<PAP>}
\eeq
where $\hat{P}$ is the projector of Eq. (\ref{P}) and $\hat{A}$ is the observable of Eq. (\ref{A}).
The expectation value (\ref{<PAP>}) apparently may be underpinned via a HV model through WFs, simply because $W_{|\psi\rangle}$ is Gaussian and the observable 
$\hat{P}\hat{A}\hat{P}$ has a proper WF. 
However, if we let $\hat{P}$ act on $|\psi\ran$ first, we obtain $|\psi'\ran$, whose WF obtains also negative values. Therefore, the HV interpretation of this average via the WF does not hold anymore! 

The above results, which were obtained with the particular projector (\ref{P}), motivate the study of more general projective measurements which we now undertake.
We know that the only pure state whose WF is non-negative is a Gaussian. Therefore, the only way to preserve the non-negativity of a WF after a projection is to obtain a new Gaussian.
Let $|\psi_{G_1}\ran$ be a state which, in the coordinate representation, 
$\lan x|\psi_{G_1}\ran = \psi_{G_1}(x)$, is a Gaussian.
Then the projector 
\beq
\hat{P}_{G_2} = |\psi_{G_2}\ran \lan \psi_{G_2}|,
\label{PG2}
\eeq
(where $|\psi_{G_2}\ran$ is such that $\lan x|\psi_{G_2}\ran $ is, in general, another Gaussian) acting on $|\psi_{G_1}\ran$ gives
\beq
\frac{\hat{P}_{G_2}|\psi_{G_1}\ran}
{\lan \psi_{G_1}| P_{G_2} |\psi_{G_1}\ran^{1/2}}
=|\psi_{G_2}\ran \; .
\label{PG2.G1}
\eeq
Thus there is an infinity of projectors that preserve the non-negativity of the WF of the original state.
However, the WFs of these projectors, i.e.,
\beq
W_{\hat{P}_{G_2}} (q,p)
=2\,e^{-\frac{q^2}{2\sigma_{G_2}^2}}\,e^{-2\sigma_{G_2}^2 p^2} \; ,
\label{W_P_G2}
\eeq
do not take on the values 1 and 0 
(the eigenvalues of the observable $\hat{P}_{G_2}$) only, and thus do not qualify as ``proper observables'' \cite{rev} in phase space.

In contrast, the WF (\ref{W_P}) of the projector $\hat{P}$ of Eq. (\ref{P}) is a proper observable in phase space; however, $\hat{P}$ is not of the form (\ref{PG2}) and thus does not preserve the non-negativity of the WF of the original state.

Starting from (\ref{PG2}) we can now build higher-rank projectors 
\beq
\hat{P}'_{G_2} = |\psi_{G_2}\ran \lan \psi_{G_2}|
+ \sum_{k=1}^N  |\psi_{k}\ran \lan \psi_{k}| \; ,
\label{PG2tilde}
\eeq
which have the same property (\ref{PG2.G1}) if the $|\psi_{k}\ran$'s are orthogonal to $|\psi_{G_1}\ran$;
we also need the $|\psi_{k}\ran$'s to be orthogonal to $|\psi_{G_2}\ran$, so that
$\hat{P}'_{G_2}$ is again a projector.
We have seen that the WF of the first term of the projector (\ref{PG2tilde}) is
not a proper one; it seems likely that the addition of further terms will not make it proper, although we have not succeeded in proving this statement.

We conclude that {\em the only projectors that preserve the non-negativity of the WF of a pure state are of the form (\ref{PG2tilde}). 
However, at least for $N=0$, the corresponding WFs 
do not qualify as ``proper observables" in phase space}.


From the above discussion it appears that the identification of (at least) phase variables as hidden variables is erroneous. 
Even for a state with a non-negative WF and an observable with a proper WF, the result of a projective measurement can no longer be described in terms of hidden variables endowed with a classical probability density.
Although for simplicity we concentrated the discussion on single-particle systems, we believe that our conclusion has a wider generality.
For instance, in the problem of continuous-variable teleportation (a three-particle problem), the so-called ``standard protocol"  \cite{bennet_et_al} considers a projective measurement as one of its basic operations. 
Thus, even within the domain of Gaussian wave functions, we conclude that using WFs one cannot describe this problem in terms of LHVs obeying classical statistics.


\end{document}